\documentclass[aps,prb,twocolumn,showpacs,preprintnumbers,amsmath,amssymb,superscriptaddress]{revtex4-1}

\usepackage{graphicx}
\usepackage{dcolumn}
\usepackage{bm}
\newcommand{\nc}{\newcommand}
\nc{\be}{\begin{equation}}
\nc{\ee}{\end{equation}}
\nc{\bea}{\begin{eqnarray}}
\nc{\eea}{\end{eqnarray}}
\nc{\bean}{\begin{eqnarray*}}
\nc{\eean}{\end{eqnarray*}}
\nc{\mb}{\mbox}
\nc{\rnc}{\renewcommand}
\nc{\vk}{\mb{\bf k}}
\nc{\vp}{\mb{\bf p}}
\nc{\vn}{\mb{\bf n}}
\nc{\vq}{\mb{\bf q}}
\nc{\rr}{\mb{\bf r}}
\nc{\vz}{\hat {\mb{\bf z}}}
\nc{\vj}{\mb{\boldmath$j$}}
\nc{\vg}{\mb{\boldmath$g$}}
\nc{\x}{\mb{\boldmath$x$}}
\nc{\A}{\mb{\boldmath$A$}}
\nc{\va}{\mb{\boldmath$a$}}
\nc{\vs}{\mb{\boldmath$\sigma$}}
\nc{\vpi}{\mb{\boldmath$\pi$}}
\nc{\nab}{\nabla}
\nc{\X}{\sf x}
\nc{\kk}{{\bf k}}
\nc{\pp}{{\bf p}}
\nc{\upspin}{{\uparrow}}
\nc{\dspin}{{\downarrow}}
\nc{\vecq}{{\bf q}}
\nc{\veck}{{\bf k}}
\nc{\vecp}{{\bf p}}
\nc{\vecl}{{\bf l}}
\nc{\vecr}{{\bf r}}
\nc{\vecx}{{\bf x}}
\nc{\vecR}{{\bf R}}
\nc{\vecG}{{\bf G}}
\nc{\vecA}{{\bf A}}
\nc{\vecpi}{{\bf \pi}}
\nc{\vecL}{{\bf L}}
\nc{\vecK}{{\bf K}}

\begin{document}

\begin{abstract}
Mirror symmetric twisted trilayer graphene (tTLG) is composed of even parity twisted bilayer graphene (tBLG)-like bands and odd parity Dirac-like bands. Here, we study the mirror-symmetric and mirror-asymmetric Hofstadter-Moir\'{e} (HM) fractal bands of tTLG. A novel quantum parity Hall state is identified in mirror-symmetric tTLG at experimentally accessible charge densities. This mirror symmetry-protected topological phase exhibits simultaneous quantized Hall and longitudinal resistances. The effects of the displacement field on the HM fractal bands of tTLG and topological phase transitions are also studied. The application of an electric displacement field results in an emergent weakly dispersive band at the charge neutrality point for a range of twist angles. This zero-energy state resides in the middle layer. It is isolated from the HM spectrum by an energy gap that scales proportional to the applied displacement field, making it a prime candidate to host correlated topological states. 
\end{abstract}

\title{Hofstadter-Moir\'{e} Butterfly in Twisted Trilayer Graphene}
	
\author{Muhammad Imran}
\affiliation{Department of Physics, University of Nevada, Reno, Nevada 89557, USA}
\author{Paul M. Haney}
\affiliation{Physical Measurement Laboratory, National Institute of Standards and Technology; Gaithersburg, MD 20899, USA}
\author{Yafis Barlas}
\affiliation{Department of Physics, University of Nevada, Reno, Nevada 89557, USA}
\date{\today}
\maketitle

Owing to the large Moir\'{e} superlattice periodicity ($a_M \sim 10- 50 $ nm), when compared to lattice spacing, twisted 2D crystals provide an ideal test bed for investigation of the various topological phases of Hofstadter-Moir\'{e}  fractal patterns~\cite{PhysRevB.14.2239,RBAM,HLB,PhysRevB.102.041402,PhysRevB.14.2239,Dean2013,PhysRevLett.49.405,PhysRevB.84.035440,PhysRevB.102.035421}. These fractals are significantly influenced by twist angles and substrate interactions~\cite{RBAM,HLB,PhysRevB.102.041402}. More importantly, electron-electron interactions in the Hofstadter-Moir\'{e} bands can result in correlated~\cite{AndreaStoner} and exotic topological phases~\cite{Xie2021}. Such ``twistronic" engineering of flat bands in twisted 2D crystals is a promising route to discover novel interaction-driven correlated and topological phases~\cite{Bistritzer12233,Cao2018,Yankowitz1059,Cao2018second, Chen2019,Lu2019,Sharpe605,CStDLG_2020,CStDLG_Pablo,CStDLG_Kim,Polshyn_2020,CherntMLG_Cory,PhysRevLett.108.216802,PhysRevLett.122.106405}. A recently discovered class of these systems is the alternating twist multilayer graphene. They consist of $m \geq 3$ graphene monolayers with a twist angle ($\theta$) that alternates between $+\theta$ and $-\theta$ between each successive pair of layers~\cite{PhysRevLett.123.026402,PhysRevB.100.085109,Cea2019,SCMAtTLGPablo,TripletSCtTLGPablo, TutuctQLG,VPF}. In alternating twisted multilayer graphene, electric fields perpendicular to the sample, in addition to twist angles and substrate interactions, can significantly modify the Hofstadter-Moir\'{e} fractal patterns. This tunability of the Hofstadter-Moir\'{e} fractals can result in emergent regimes that might be ideal for realizing novel correlated and topological phases~\cite{AndreaStoner,Xie2021}.

In this paper, we report on the topological properties and energy bands of Hofstadter-Moir\'{e} fractals in twisted trilayer graphene (tTLG).
Without a displacement field, tTLG obeys mirror symmetry~\cite{PhysRevX.2.011004,LSHL,SMAA,PhysRevLett.121.066602}. This allows for decomposition into tBLG-like even-parity bands and monolayer graphene (MLG)-like Dirac odd-parity bands~\cite{PhysRevLett.123.026402,PhysRevB.100.085109,VPF}. At high magnetic fields, this results in the co-existence of a tBLG-like Hofstadter-Moir\'{e} pattern with an MLG-like Landau level (LL) spectra in mirror-symmetric tTLG. Our results for the even parity tTLG Hofstadter-Moir\'{e}  patterns at zero displacement field are consistent with earlier studies of the Hofstadter-Moir\'{e}  patterns in tBLG~\cite{RBAM,HLB,PhysRevB.102.041402}. However, tTLG Hofstadter-Moir\'{e}  patterns exhibit a different sequence of Chern numbers due to the simultaneous presence of the odd parity MLG-like LLs. 

\begin{figure}[t]
\begin{center}
	\includegraphics[width=0.45\textwidth,clip]{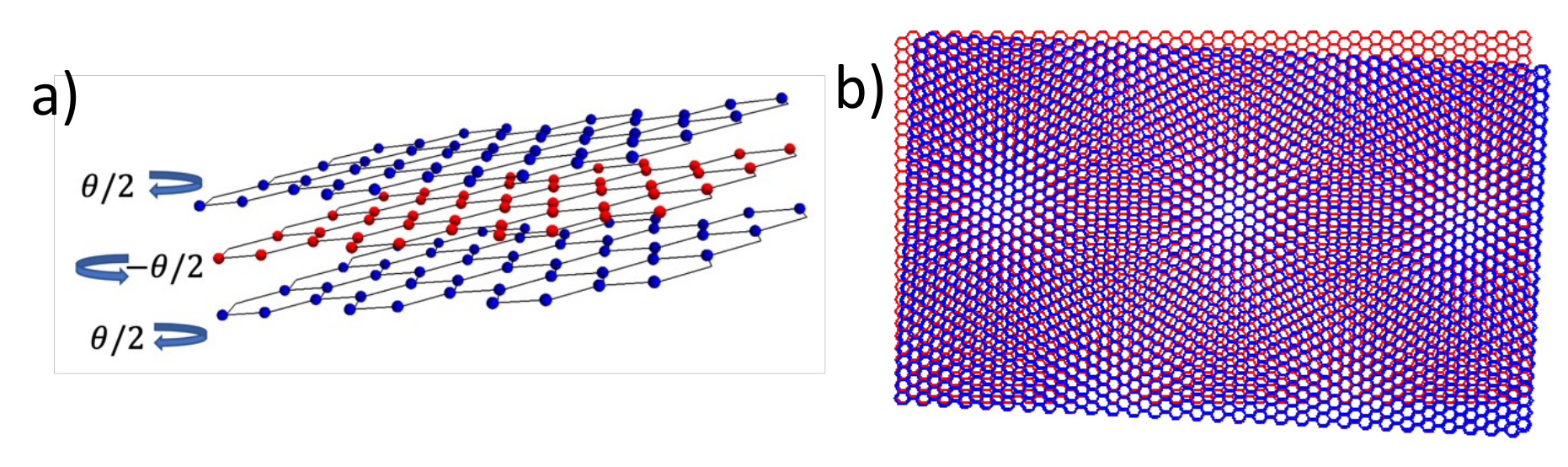}	
		\caption{ a) Lattice structure of twisted trilayer graphene with the top and bottom layer rotated by $\theta/2$ while the middle layer is rotated by 
$-\theta/2$. b) Moir\'{e} pattern in twisted trilayer graphene with local AA and AB stacking regions.}
\label{fig:Fig1tTLG}
\end{center}
\end{figure}

More importantly, the mirror symmetry stabilizes a symmetry-protected topological phase, which we call the quantum parity Hall state. It originates from unequal counter-propagating branches of even and odd parity edges due to unequal and opposite signs of the Hall conductivity in each parity sector. This state is present in the angle regimes $\theta \approx 1.6^{\circ}$ to $2.5^{\circ}$ at accessible charge densities in tTLG. Since mirror symmetry forbids backscattering between different parity sectors, this state {\it exhibits simultaneous quantization of the Hall and longitudinal resistances}. Similar quantum Hall parity states have been identified in ABA-stacked trilayer graphene at neutral charge density~\cite{PhysRevLett.121.066602,Stepanov10286}.  

\begin{figure*}[th]
\begin{center}
\includegraphics[width=0.95\textwidth,clip]{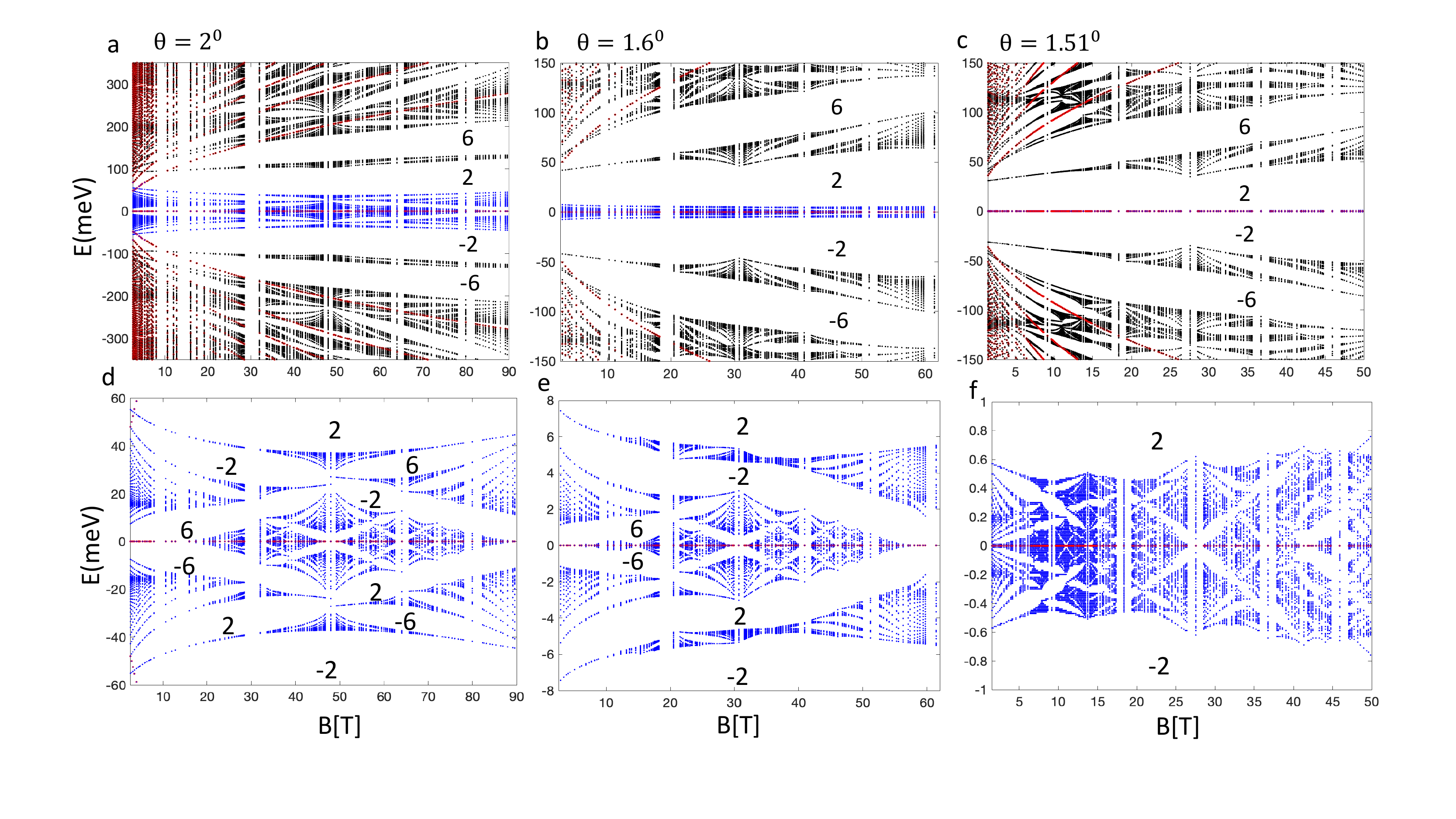}	
\caption{Hofstadter-Moir\'{e} butterfly patterns in tTLG for zero displacement field. Energy eigenvalue dispersion as a function of the applied magnetic field at the angles a) $\theta = 2^{\circ} $,  b)  $\theta=1.6^{\circ}$ and c) $\theta = 1.51^{\circ} $(magic angle) with $w = 97.50 ~{\rm meV}$ and $\eta = 0.82 $. In a), b), and c), the even parity bands are denoted by black/blue, while the odd parity bands are denoted by red. The values in the spectral gaps indicate the Hall conductivity $\sigma_{xy}$ in units of $e^2/h$. For clarity, the details of the central band regions in a), b), and c), depicted in blue for even parity bands and red for the odd parity bands, are magnified in d), e), and f).}
\label{fig:tTLGHMbands}
\end{center}
\end{figure*}

In a displacement field, the Hofstadter-Moir\'{e}  pattern is significantly modified due to the hybridization of the tBLG-like band with the MLG-like Landau levels (LLs). The system exhibits a fractured Hofstadter-Moir\'{e}  butterfly pattern, followed by an emergent zero-energy weakly dispersive band. This weakly dispersing flat band is pinned to the charge neutrality point and separated from the rest of the spectrum by a band gap. In the angle regime $\theta \approx 1.7^{\circ}$ to $2.5^{\circ}$, the band gap increases linearly with the applied displacement field energy $\Delta_{\perp}$. This is accompanied by a slight increase in the bandwidth for the range $\Delta_{\perp} = 5~{\rm meV}$ to $30~{\rm meV}$. 
This zero-energy band resides primarily in the middle layer. Its energetic and topological properties are tunable by the applied displacement field, making it a promising candidate for hosting many-body interacting ground states~\cite{Xie2021}.

The paper is organized as follows. In section A, we discuss the mirror-symmetric Hofstadter-Moir\'{e} butterfly, its band dispersion, and topological properties as a function of twist angles. Section B studies the origin and properties of the emergent zero-energy weakly dispersive band induced by electric fields in tTLG. Finally, in section C, we discuss the relevance of our results to experiments on tTLG at high magnetic fields.  The details of the calculations, model Hamiltonian of tTLG, and methods are relegated to the Appendices.	
	
\subsection{Mirror symmetric tTLG Hofstadter-Moir\'{e} butterfly}

The lattice structure of the tTLG lattice exhibits mirror symmetry about the middle layer, as indicated in Fig.~\ref{fig:Fig1tTLG} (a). This facilitates a description of energy bands in parity eigenstates~\cite{LSHL,SMAA,VPF}. We denote sublattice $A(B)$ on layer $i$ with $A_i(B_i)$.  The even parity orbital combinations are then given by $(A_+,B_+,A_2,B_2)$ while the odd parity orbitals are $(A_-,B_-)$, where $A_{\pm} = (A_{1} \pm A_{3})/\sqrt{2}$, and a similar expression applies for $B_{\pm}$. We take the relative in-plane displacement, $d=0$, and denote the top (bottom) layers angle $\theta/2$, while the middle layer angle, $-\theta/2$. The band dispersion due to the Mori\'{e} pattern formed at small twist angles can be captured by extensions of the Bistritzer-MacDonald (BM) Hamiltonian~\cite{Bistritzer12233}. The BM model captures the effect of the periodic tunneling between the layers in the AA and AB stacked regions (see Fig.~\ref{fig:Fig1tTLG} (b)), denoted by $w_{AB} = w = 97.5~ {\rm meV}$ and $w_{AA} = \eta w $ with $\eta = 0.82$, respectively (see Appendix A for the tTLG Hamiltonian). At zero displacement fields, due to mirror symmetry, the Hamiltonian can be decomposed into tBLG-like Hamiltonian with enhanced tunneling parameter $ w \to \sqrt{2} w $, and an MLG-like Dirac band (see Appendix A)~\cite{VPF,PhysRevB.100.085109}. 

The large Moir\'{e} periodicity of twisted 2D crystal results in fractal Hofstadter-Moir\'{e} (HM) bands at high magnetic fields. We used the parity eigenstate basis to calculate the HM-bands of tTLG with the gauge choice ${\bf A} = B(-y,0)$. The Hamiltonian was expressed in the basis set, $\{ | n, Y_i ,\alpha , \sigma \rangle \}$, where $n$ denotes the Landau level (LL) index at the guiding center positioned at $Y_{i}$, (which corresponds to a lattice site in the unit cell) on the sublattice $\alpha $. The index $\sigma = 1,2,3$ denotes the even(odd) parity eigenspinors with the assignments $1=(A_{+},B_{-})$, $2=(A_2,B_{2})$ and $3=(A_-,B_-)$. The details of the calculation are presented in Appendix A.

\begin{figure}
\begin{center}
\includegraphics[width=0.45\textwidth,clip]{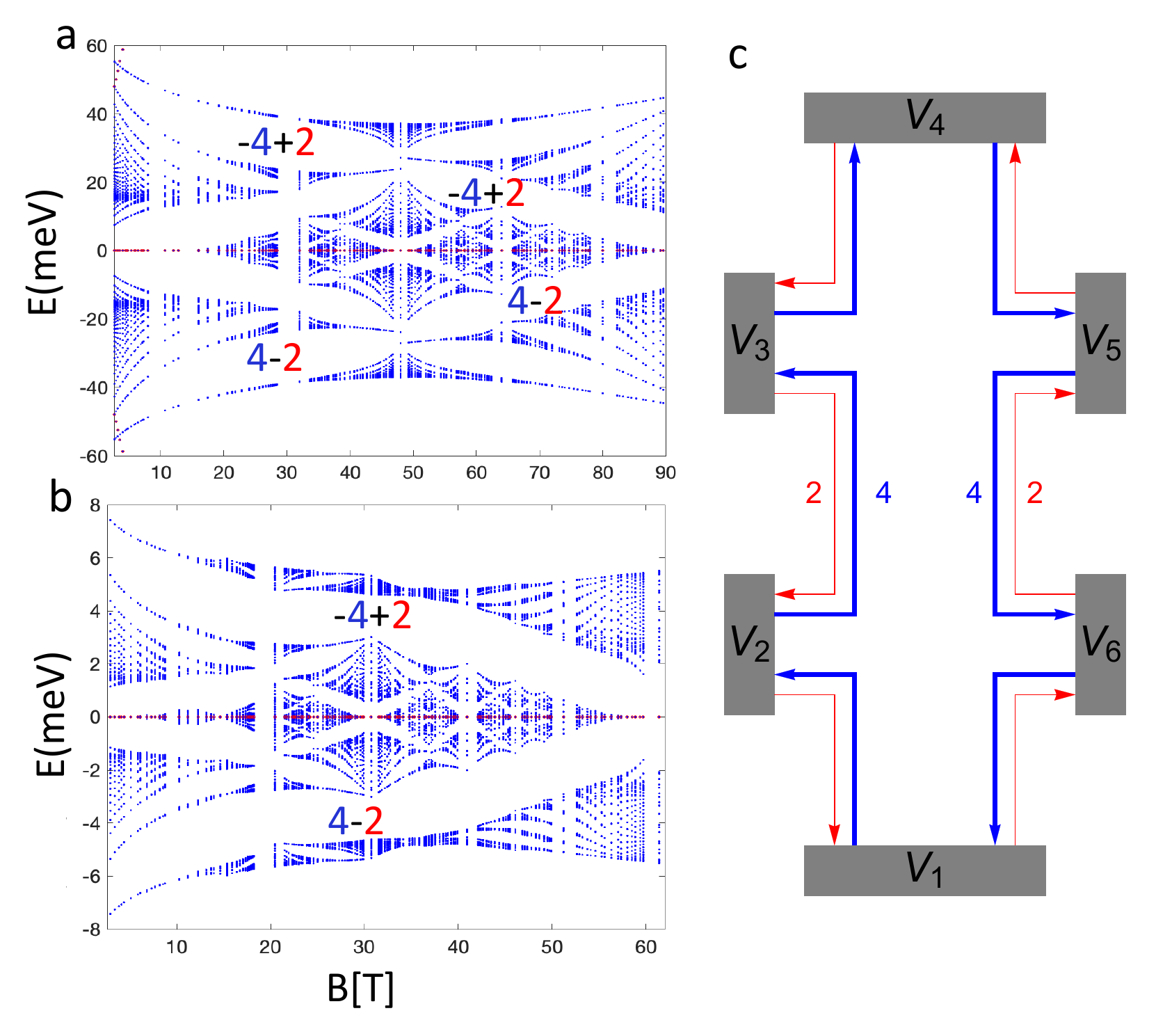}	
\caption{Quantum parity Hall phase in the Hofstadter-Moir\'{e} butterfly patterns of tTLG for a) $\theta = 2^{\circ} $,  b)  $\theta=1.6^{\circ}$. The number of edge states associated with the even-parity tBLG-like bands is in blue, while the MLG-like bands are in red, in units of $e^2/h$. c) Edge state schematic of the quantum parity Hall phase in a six-terminal Hall-bar geometry at positive charge densities.}
\label{fig:QPHtTLG}
\end{center}
\end{figure}

Our calculations for tTLG exhibited rich structures in the HM spectrum, which can be tuned by the electric field and twist angles. Fig.~\ref{fig:tTLGHMbands} shows the HM butterfly for mirror-symmetric tTLG at three representative angles ($\theta = 2^{\circ}, 1.6^{\circ}$ and $1.5^{\circ}$). The Hall conductivity, in units of $e^2/h$, is shown in the spectral gaps. In Fig.~\ref{fig:tTLGHMbands}, the even parity bands are depicted in blue or black, while the odd parity bands are shown in red. The Landau bands originating from the odd parity sector can be distinguished by $\epsilon_n \propto \sqrt{B}$, while the energy of the even parity bands shows a tBLG HM fractal pattern. Similar HM butterfly patterns for tBLG have been reported in Ref.~\onlinecite{HLB}. Our even parity band HM butterfly patterns are consistent with these reports but now occur at twice the magnetic fields due to the $\sqrt{2}$ enhancement of the twist angle in the tTLG even parity sector. 

\begin{figure*}
\begin{center}
\includegraphics[width=0.95\textwidth,clip]{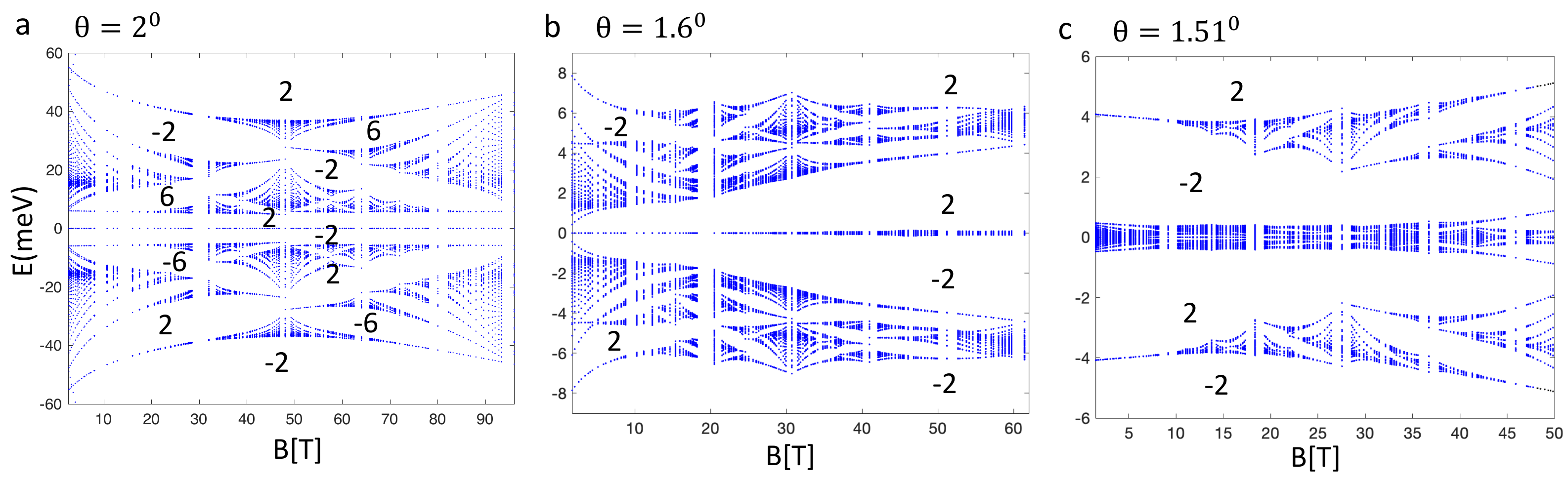}	
\caption{Hofstadter-Moir\'{e} butterfly patterns in tTLG for an electric displacement field strength $ \Delta_{\perp} =10 ~{\rm meV}$ at the angles a) $\theta = 2^{\circ} $,  b)  $\theta=1.6^{\circ}$ and c) $\theta = 1.51^{\circ} $(magic angle) with $w = 97.50~ {\rm meV}$ and $\eta = 0.82 $. The values in the spectral gaps indicate the Hall conductivity $\sigma_{xy}$ in units of $e^2/h$. }
\label{fig:tTLGHMEfield}
\end{center}
\end{figure*} 

We primarily focused on three angles, each indicating three distinct regimes of the HM butterfly. A detailed version of these central bands of the HM butterfly is shown in Fig.~\ref{fig:tTLGHMbands} d), e), and f). The $\theta = 2^{\circ}$ HM butterfly is representative of the twist angle range $\theta \approx 1.7^{\circ}$ to $2.5^{\circ}$. In this regime, we found an emergent Hofstadter pattern similar to the Hofstadter pattern of the tight-binding model for graphene. In contrast, for $\theta = 1.6^{\circ}$, we found a spectral gap for all magnetic fields. Similar results were obtained for the range of angles $\theta \approx 1.65^{\circ}$ to $1.55^{\circ}$, after which the pattern changed significantly.  At the magic angle $\theta =1.51^{\circ}$, the HM pattern is modified with no resemblance to the Hofstadter pattern in monolayer graphene. The bandwidth of the central bands decreases nearly an order of magnitude compared to the HM pattern at $\theta = 2^{\circ}$. Below the magic angle at $\theta=1.45^{\circ}$ another pattern reemerged similar to $\theta = 1.6^{\circ}$.  

In Fig.~\ref{fig:tTLGHMbands}, the integers in the spectral gaps of the HM butterflies denote the Hall conductivity, $\sigma_{H}$ in units of $e^{2}/h$. The numerically attained eigenfunctions were employed with the Wilson loop procedure~\cite{Chernnumbercalculation} to calculate the Chern numbers and Berry flux (see Appendix B for details of this method). We calculated the Hall conductivity within the larger spectral gaps $\approx 5 \geqslant \rm{meV}$. The Hall conductivity at the charge neutrality point $\sigma_{H} (\epsilon_{F} = 0 ) = 0$ was regularized to zero and included the spin and valley degeneracy. The Chern numbers and Hall conductivity of the emergent HM pattern for $\theta = 2^{\circ}$ in the even parity sector of tTLG are the same as the monolayer graphene Hofstadter butterfly. This aspect of the duality for tBLG has been reported in Ref.~\onlinecite{HLB}. However, in tTLG, the Hall conductivity is the sum of the Hall conductivity of tBLG even-parity HM bands and the MLG-bands odd parity LLs. 

A consequence of mirror-symmetry in tTLG is a symmetry-protected topological (SPT) phase that simultaneously quantizes longitudinal and Hall resistance. This mirror-SPT (mSPT) phase, which we call the quantum parity Hall phase, was identified at neutral charge density in ABA trilayer graphene~\cite{PhysRevLett.121.066602,Stepanov10286}. In tTLG, this state occurs at finite charge density. It is marked by unequal branches of counterpropagating even-parity and odd-parity edge modes associated with tBLG-like HM bands and MLG-like LL bands. In Fig.~\ref{fig:QPHtTLG} a) and b), we label the regions where the quantum parity Hall state appears by the number of edge states associated with each parity sector, blue(red) for even(odd)-parity. 

In these regions, the Hall conductivity is positive(negative) for negative(positive) energies. Since the neutral charge density is defined at zero energy, this corresponds to a positive(negative) sign of Hall conductance for hole-like(electron-like) charge densities. This is an essential feature of this quantum Hall parity state in tTLG. From our calculations of Chern numbers, we only found one instance of this state. Still, other types of mSPT phases can be realized in regions with smaller spectral gaps $\lessapprox 5 ~\rm{meV}.$ They can be identified by negative(positive) even-parity tBLG-bands Chern numbers at positive(negative) charge densities.

Fig.~\ref{fig:QPHtTLG} c) shows the edge states for the quantum parity Hall phase in a six-terminal Hall bar geometry for positive charge densities. There are  $2$ edge modes originating from the odd parity LL bands (shown in red) and $4$ counter-propagating edge modes arising from the even-parity Hofstadter bands (shown in blue). Since the edge states in the mirror sectors have unequal branches of edge modes, they exhibit simultaneous quantized Hall and longitudinal resistances. The edge modes are protected from back-scattering by mirror symmetry. The resistances in the Hall bar geometry can be calculated from the Landauer-Buttiker approach~\cite{PhysRevB.38.9375} (see Appendix C) for the quantum parity Hall state, giving
\be
R_{14,26} = \frac{h}{6e^2} ; \qquad R_{14,32} = \frac{h}{9e^2} ; \qquad R_{14,14} = \frac{4h}{9 e^2},
 \ee
where $R_{ij,kl}$ is defined as the ratio of the voltage to the current measured between the $k^{th}$ and the $j^{th}$, with current applied from the $i^{th}$ to the $j^{th}$ lead.
The edge states of the quantum parity Hall phase and their stability to disorder will be discussed elsewhere. Next, we study the effect of the displacement field on the HM patterns on tTLG.

\subsection{Emergent zero-energy state in tTLG}
	
The displacement field breaks mirror symmetry, hybridizing the Dirac LLs with the even parity HM bands of the tBLG-like even parity sector. Fig.~\ref{fig:tTLGHMEfield} $(a), (b)$ \& $(c)$ shows the HM pattern in the presence of a displacement field of strength $\Delta_{\perp} = 10$ meV. The most striking feature in Fig.~\ref{fig:tTLGHMEfield} $(a), (b)$ \& $(c)$ is the emergence of two spectral gaps adjacent to the charge neutrality point. For all three angle regimes, we observed a fractured fractal pattern in a displacement field when compared to the HM fractal patterns in Fig.~\ref{fig:tTLGHMbands} d), e), and f). This is accompanied by the emergence of a weakly dispersing zero-energy band pinned at the charge neutrality point. This zero-energy band disperses with a small bandwidth $\approx 0.1~{\rm meV}$ to $0.2  $ meV for the twist angle $\theta=2^{\circ}$. However, its bandwidth slightly increases at smaller twist angles as a function of the magnetic field. The spectral gap at zero-energy is given by $\approx \Delta_{\perp}/2$ for $\theta= 2^{\circ} $, and it is independent of the magnetic field within numerical accuracy. This spectral gap results from a level repulsion mechanism, as discussed below.  

Another striking feature is the change of the Hall conductance as a function of the twist angle. The Hall conductance at $\theta = 2^{\circ}, 1.6^{\circ}$ adjacent to the zero-energy state is $\sigma_{xy}= -2e^2/h$ and $2e^2/h$ changes to $\sigma_{xy}= 2e^2/h$ and $-2e^2/h$ at $\theta =1.51^{\circ}$ as a function of the twist angle. This topological transition indicates a significant band reconstruction between the twist angles $\theta = 1.6^{\circ}$ and $\theta = 1.51^{\circ}$. These transitions are associated with the Berry curvature's tunability and band dispersion as a function of the electric field. This phase transition is evident in the corresponding Wannier plots for tTLG (see Appendix B).     
 
\begin{figure}
	\begin{center}
		\includegraphics[width=0.45\textwidth,clip]{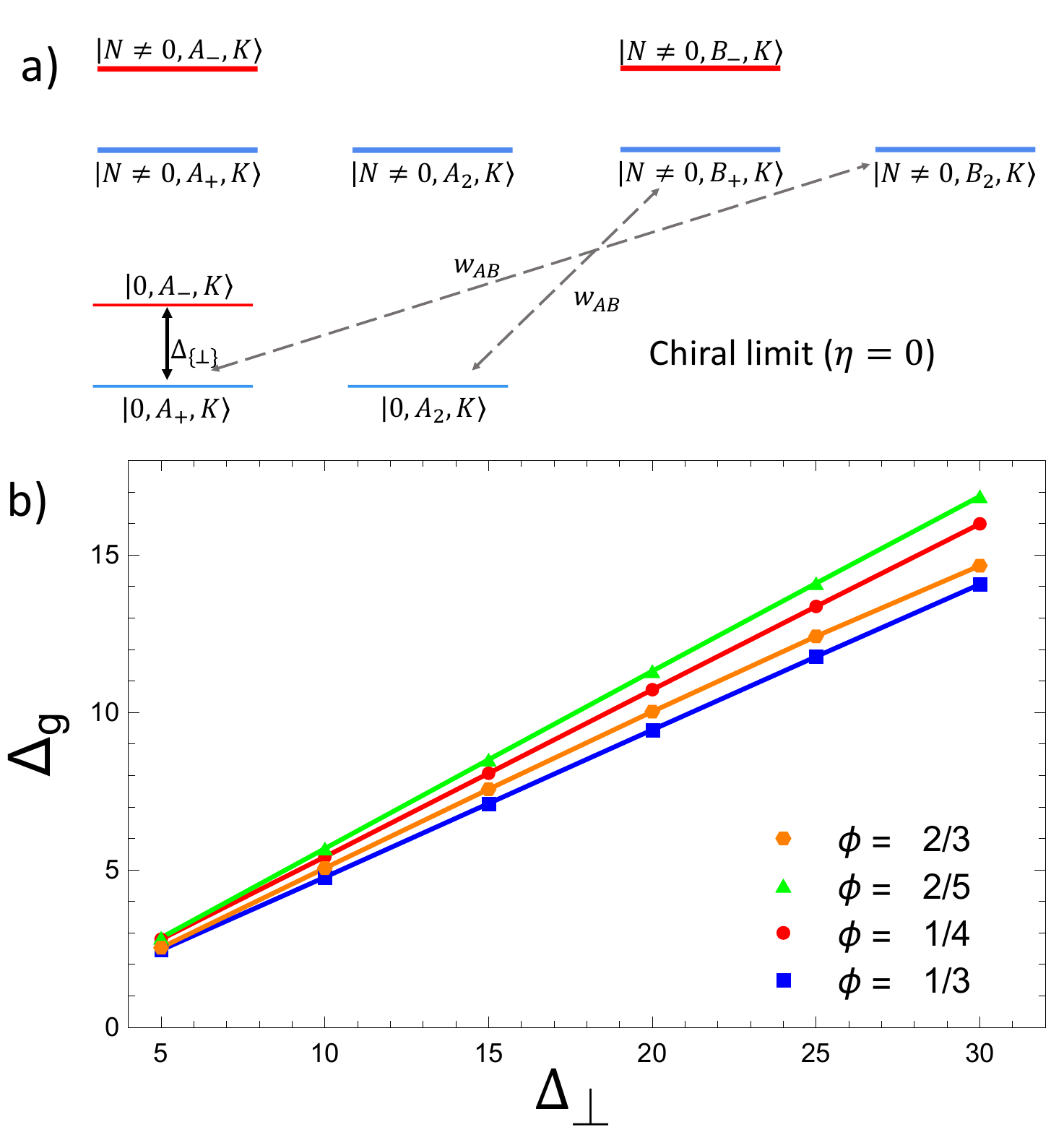}
		\caption{a) Schematic representation of the coupling in the chiral limit ($\eta =0$) in valley ${\bf K}$ for the $N=0$ with the $N \neq 0$ LLs. The direct coupling in $N=0$ LL due to the displacement field is represented by the solid black line, while the perturbative coupling due to tunneling between the $A$ and $B$ sublattices is denoted by the dashed line. b) Band gap as a function of the displacement field for different values of $\phi = p/q$, all energies are in meV. All results are for the twist angle $\theta = 2^{\circ}$. }	
		\label{fig:Zeneprop}
	\end{center}
\end{figure}

The most striking feature is the emergence of a zero-energy flat band multiplet in the angle regimes $\theta = 1.7^{\circ}$ to $2.5^{\circ}$ under a displacement field. This zero-energy flat band multiplet is q-fold degenerate and completely resides in the middle layer. The origin of the zero-energy band can be understood by starting in the chiral limit ($\eta=0$) and zero displacement field and projecting on the $N=0$ LL index. The chiral limit, defined by $\eta =0$, corresponds to the absence of tunneling between the same orbitals (i.e. $w_{A_+ A_2}=w_{B_+ B_2} =0$) in the even parity tBLG-like bands~\cite{AshwinTMLG}. Therefore, in a magnetic field, when $\eta=0$, the $N=0$ LL in valley ${\bf K}$ lies on the sublattice $A_{+}, A_{2}$ in the even parity sector, and  $A_{-}$ in the odd parity sector, while in valley ${\bf K'}$ the zeroth LL lies on the sublattice $B_{+}, B_{2}$ in the even parity sector, and  $B_{-}$ in the odd parity sector.  Since the $N=0$ LLs are localized on the sublattice, there is no direct coupling between the $N=0$ LLs, as indicated in Fig.~\ref{fig:Zeneprop} a).  However, the $N \neq 0 $ LL are perturbatively coupled to the $N=0$ LL  due to $w_{A_+ B_2}$ and $w_{B_+ A_2}$ tunneling. In Fig.~\ref{fig:Zeneprop} a), this mixing is indicated by the dashed lines, where we only show the coupling in valley ${\bf K}$. Below, we present the argument for the level mechanism for the ${\bf{K}}$ valley, the other valley ${\bf{K'}}$ can be attained by interchanging the sublattices.
 
When a displacement field is applied, the $A_+$ orbital hybridizes with the $A_{-}$ orbital in the odd-parity sector. This direct coupling is shown in the solid line in Fig.~\ref{fig:Zeneprop} a). The hybridization induced by the displacement field couples the $N=0$ LLs on the sublattice $A_{+}$ with $A_{-}$ in the ${\bf K}$ valley. These states gap out due to level repulsion, leaving the zero-energy state on the middle layer on the orbitals $A_{2}$ in ${\bf K}$ at zero energy. Therefore, in the chiral limit $\eta =0$, the emergent zero-energy state in the HM pattern at $\theta =2^{\circ}$ is localized in the middle layer on sublattice $A_{2}$ in the ${\bf K}$ valley. It is essential to point out that this zero-energy level has components $N \neq 0 $ LL due to mixing induced by the  $w_{A_+ B_2}$ and $w_{A_2 B_+} $ tunneling terms. In the chiral limit, the calculated projected weight of the emergent zero-energy state averaged over the BZ-mesh on the $N=0$ LL orbital in the middle layer was $\approx 0.8$, indicating mixing with higher LL int the middle layer. 

When $\eta =0.82 $ the $N=0$ LL in the middle layer is weakly coupled to the $N \neq 0$ LL by a combination of the displacement field and the tunneling terms $w_{A_+ A_2}\neq 0$, and  $w_{B_+ B_2}  \neq 0$. However, even when $\eta =0.82$, we found that the emergent zero-energy state entirely resided in the middle layer. We verified our analysis by projecting the wavefunction amplitude of the emergent zero-energy state on the middle layer. The projected amplitude of the emergent zero-energy state averaged over the BZ-mesh, and the multiplet band index on the middle layer came out to be $\approx 1$. This projected amplitude was calculated for different values $ \phi$. The same results are obtained for various displacement fields. Furthermore, the calculated projected weight of the emergent zero-energy state averaged over the BZ-mesh on the $N=0$ LL orbital in the middle layer was $\approx 0.6$, indicating significant mixing with $N\neq 0$ LLs in the middle layer. 

More evidence of the level repulsion mechanism can be inferred from the behavior of the energy gap above the zero-energy state $ \Delta_g$ as a function of the displacement field. The energy gap $ \Delta_g \approx \Delta_{\perp}/2 $ grows linearly as a function of the displacement field (see Fig.~\ref{fig:Zeneprop} b). The bandwidth of the zero-energy state $\Delta_w \approx 0.1~{\rm meV}$ to $0.4 $ meV is much smaller than the bandgap and varies slightly with the electric field. We also found that the Berry curvature deviation of the zero-energy state decreases as a function of the displacement field strength $\Delta_{\perp}$. This tunability of the Berry curvature and isolation of the emergent zero-energy state provide ideal conditions for realizing various interesting many-body interacting ground states~\cite{Xie2021}. Since the emergent zero-energy state has significant mixing of higher LL wavefunction, it is anticipated that the ground state at fractional filling will most likely be a Wigner crystal or charge density wave state~\cite{PhysRevLett.76.499,PhysRevB.54.1853,PhysRevB.55.9326,PhysRevLett.82.394,PhysRevLett.109.126804}. Due to the complexity of the computational basis, these studies must be performed on lattice analogs of the HM pattern of tTLG.

\subsection{Conclusion and Outlook}
The displacement field provides an external knob to manipulate the topological phase and energy spectrum in the HM butterfly. The most striking is the emergence of a zero-energy state at the charge neutrality point within an accessible range of doping densities, whose separation from the energy spectrum can be tuned by the displacement field. The narrow bandwidth of the zero-energy band indicates the possibility of strongly correlated phases such as quantum Hall ferromagnetism~\cite{PhysRevLett.96.256602,PhysRevLett.109.046803,PhysRevLett.117.076807,Barlas_2012}, possible charge density waves~\cite{PhysRevLett.76.499,PhysRevB.54.1853,PhysRevB.55.9326,PhysRevLett.82.394,PhysRevLett.109.126804}, and fractional topological insulators~\cite{Xie2021,PhysRevB.85.241308,PhysRevX.1.021014,PhysRevB.101.235312,PhysRevResearch.2.023238,PhysRevResearch.2.023237}. Furthermore, the electric field can be used to access topological transitions. This makes it possible to probe the HM butterfly patterns in tTLG in transport or via scanning experiments. In addition, we discovered a symmetry-protected topological phase for the mirror-symmetric case due to unequal counter-propagating edge modes exhibiting simultaneous Hall and longitudinal resistances. Interactions within each sector of the quantum parity Hall phase will most likely result in an analog exotic correlated quantum Hall phases detected in the ABA stacked trilayer graphene~\cite{PhysRevLett.121.066602,Stepanov10286}.

\acknowledgments{M. I. and Y. B. authors acknowledge the support of UNR/VPRI startup grant PG19012. Y. B. acknowledges support from the Aspen Center for Physics, which is supported by NSF grant PHY-1607611, where part of this work was performed. }	


%

\subsection{Appendix A: tTLG Hamiltonian}

The continuum Hamiltonian of the tTLG~\cite{Bistritzer12233}, which is valid for small angles, $\theta \approx 3^{\circ}$, can be expressed in terms of a six-component spinor, 
$\psi^{\dagger}_{\bf K} = (\phi^{\dagger}_{1,\vecK}, \phi^{\dagger}_{2,\vecK},\phi^{\dagger}_{3,\vecK})$, at the Dirac point $\vecK$  
\be
H=\left( \begin{array}{ccc}
h_{\theta} + \Delta_1 & T(\vecr) & 0\\
T^{\dagger} (\vecr) & h_{-\theta} + \Delta_2 & T^{\dagger}(\vecr)\\
0 & T(\vecr) & h_{\theta} + \Delta_3
\end{array}\right),
\ee
where, $h_{\pm\theta}=\hbar v(\xi\sigma_{x},\sigma_{y}) (\veck^{\pm\theta/2}-{\vecK_{\xi}})$, denotes the Dirac Hamiltonian on the rotated Brillouin zone (BZ), $\sigma_{i}$, denotes the Pauli matrix acting on the sublattice degree of freedom,  $\xi = \pm1 $, denotes the Dirac points which correspond to different valleys at the BZ momentum $\vecK_{\xi}=4\pi/3a (\xi,0)$. The momentum space tunnelling matrix elements, $T(\vecr) = \sum_{n=1}^3  \hat{T}_{n} e^{\imath \vecq_n \cdot \vecr}$, can be expressed in terms the matrices, $\hat{T}_n$, with, 
\begin{equation}
\hat{T}_{n} = w (\eta \hat{\mathbb{I}} + \cos((n-1)\phi)) \hat{\sigma}_{x} + \sin((n-1)\phi) \hat{\sigma}_y)
\end{equation}
where, $\phi=2\pi/3$, and the tunnelling parameters are, $w_{AB} = w =97.5 $ meV, $w_{AA} = \eta w $ with $\eta =0.82$. The tunnelling matrices are related by $C_{3z}$ symmetry of the lattice via unitary operator, $U(\phi)=\exp(i\phi\sigma_{z})$. The momentum transfer vectors associated to the honeycomb Moir\'{e} lattice $\vecq_{1}=k_{\theta}(0,-1)$, $\vecq_{2}=k_{\theta}(\sqrt{3}/2,1/2)$, $\vecq_{3}=k_{\theta}(\sqrt{3}/2,-1/2)$, where $k_{\theta}=4\pi/(3a_M)$ is the distance between the mini-Dirac points and, $a_M = a/(2 \sin(\theta/2))$, is the Moir\'{e} lattice spacing. 

In the parity basis, the tTLG Hamiltonian becomes,
\be
\label{tTLGevenodd}
H(w,\eta,\Delta_{\pm})=\left( \begin{array}{ccc}
h_{\theta} + \Delta_+ & \sqrt{2} T(\vecr) & \Delta_-\\
\sqrt{2}T^{\dagger} (\vecr) & h_{-\theta} + \Delta_2 & 0\\
\Delta_-& 0 & h_{\theta} +\Delta_+ \end{array}\right),
\ee
where, $\Delta_{\pm} = (\Delta_1 \pm \Delta_3)/2$ and from now on we take $\Delta_2 =0$. The above Hamiltonian in Eq.~\ref{tTLGevenodd} is expressed in terms of a six-component spinor basis $ (A_+,B_+, A_2,B_2, A_-, B_-)$, where $A_{\pm} = (A_{1} \pm A_{3})/\sqrt{2} $ and $B_{\pm} = (B_{1} \pm B_{3})/\sqrt{2}$,  have even $(+)$ and odd $(-)$ parity with respect to this mirror symmetry, while the middle layer, $A_{2}$, and, $B_2$, orbitals have even$(+)$ parity.

\begin{figure}
\begin{center}
	\includegraphics[width=0.45\textwidth,clip]{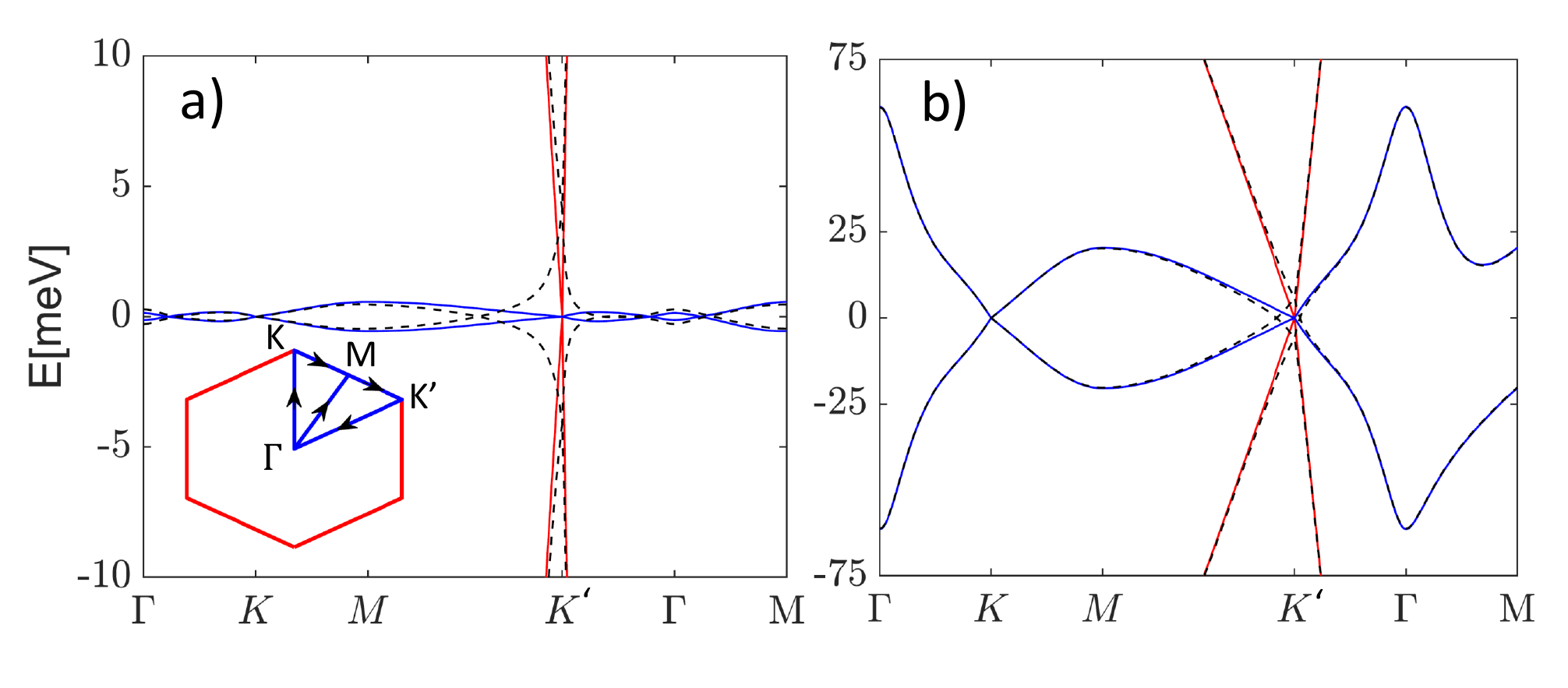}	
		\caption{ Mirror symmetric bands of tTLG at the magic angle a) $\theta = 1.51^{\circ}$ and b) $\theta = 2^{\circ}$ with even parity tBLG-like bands in blue and the monolayer-Dirac like bands in red at $ \Delta_- =0 $. The dashed line corresponds to the same angles with mirror symmetry broken ($\Delta_- = 10$ meV) by the presence of the electric field.}
\label{fig:Fig1tTLGAppendix}
\end{center}
\end{figure}

Due to the enhanced tunneling in the tBLG-like sector, the first magic angle occurs at $\theta_{M} = 1.51^{\circ} = 1.05^{\circ}\sqrt{2} $, with a bandwidth $ \approx 0.5 $ meV. Fig.~\ref{fig:Fig1tTLGAppendix} (a) and (b) shows the energy bands of tTLG at the first magic angle $\theta = 1.51^{\circ}$ and $\theta = 2^{\circ}$. In Fig. \ref{fig:Fig1tTLGAppendix} (a) and (b), the even parity energy bands are plotted in blue, and the odd parity energy bands are plotted in red. The odd parity band exhibits a Dirac-like dispersion, while the even parity bands exhibit the energy dispersion of tBLG. When the mirror symmetry is broken, for instance, by applying a displacement field, $\Delta_{\perp} =10~ {\rm meV}$ the even and odd parity bands hybridize, as indicated by the black dashed line in Fig~\ref{fig:Fig1tTLGAppendix} a) and b). 

To calculate the HM-bands of tTLG, we worked in the parity eigenstate basis, and the Landau gauge, ${\bf A} = B(-y,0)$, with the basis choice, $\{ | n, Y_i ,\alpha , \sigma \rangle \}$, where $n$ denotes the Landau level (LL) index at the guiding center positioned at, $Y_{i}$, (which corresponds to a lattice site in the unit cell) on the sublattice $\alpha $. The index $\sigma = 1,2,3$ denotes the even(odd) parity eigenspinors with the assignments $1=(A_{+},B_{-})$, $2=(A_2,B_{2})$ and $3=(A_-,B_-)$. 

The Moir\'{e} hopping pattern determines the Hamiltonian periodicity as opposed to the Moir\'{e} unit cell~\cite{RBAM}. The Moir\'{e} hopping pattern has a larger periodicity, exactly six times the periodicity of the Moir\'{e} unit cell, $ A_{Mh} = 3\sqrt{3} a_M^2$~\cite{RBAM}, where $a_{M} \approxeq a_0/\theta$ for small angles, with $a_{0} =0.142$ nm.The HM bands are calculated for rational values of flux per unit cell; with our choice of the unit cell, the flux per unit cell is given by $\phi = 3 \sqrt{3} a_M^2/(2 \pi l_B^2)= q/p$, where $p$ and, $q$ are co-primes. The magnetic field is $B=4 B_0\theta^2/\phi$, where $B_0=1~{\rm T}$ and $\theta$ is expressed in degrees. With this parametrization, the magnetic BZ (mBZ) is given by  $k_1 \in [0, 6 \pi/(\sqrt{3}\phi) ) $ and $ k_2 \in [ 0, 2 \pi/q )$ in units of $\Delta  = 3 p/ (q a_{M})$.

To achieve convergence, our LL cutoff was determined by $N = 20 ({\rm max}(w,\hbar v k_{\theta})/\epsilon_0)^2$, where $\epsilon_0 = \sqrt{2} \hbar v/l_{B}$ with $\hbar v \approx 654 ~{\rm  meV\cdot nm}$. The dimension of the matrix for tTLG graphene for a given value of $\phi$ per valley is $N_{dim} = 6q(N+1/2)$. The tTLG Hamiltonian was diagonalized over a $10 \times 10$ discrete mBZ mesh to calculate topological properties and a smaller mesh to generate the HM butterfly patterns. 

The matrix elements in the basis set can be calculated from the Hamiltonian, $H$. The matrix element associated to the diagonal Hamiltonian defined as $H_{0} = H(0,0,0)$,  are given by, 
\be
\langle n,Y_i,A,\sigma | H_0 | m, Y_j, B, \sigma' \rangle = \epsilon_0 \sqrt{n} \delta_{n,m+1} \delta_{ij} \delta_{\sigma \sigma'}, 
\ee
where, $\epsilon_0 = \sqrt{2} \hbar v/l_{B}$. It is important to point out that since the $n^{th}$ LL on sublattice A couples to the $(n+1)^{th}$ LL on sublattice B, for a finite LL cutoff $N$, the $N^{th} $ LL on sublattice A shows up at zero energy, due to numerical truncation of the Hilbert space. To circumvent this issue, we used an asymmetric cutoff in our calculations and included the LL orbitals, $n=0, \cdots, N-1 $, on sublattice A and $n=0, \cdots, N $, on sublattice B. Of course, the situation is reversed in the other valley, and the Hamiltonian is just the transpose of the Hamiltonian in the valley, ${\bf K}$.
	
The tunneling matrix elements are only non-zero for LL wavefunctions between the spinors $\sigma = 1,2$. Furthermore, due to the spatial dependence of the tunneling matrix elements, different guiding centers, $Y_j$ are coupled with different $\sigma$ indices. The tunneling matrix elements $T(\vecr)$ in the LL basis can then be expressed as,
\bea 
\nonumber
T(\vecr)  &=& \sum_{j,\veck} \sum_{n,m} \bigg[ \hat{T}_{1;\alpha, \beta} \Gamma_{1;nm} (j,\veck)
|n,Y_j,\alpha,1 \rangle \langle m,Y_j,\beta,2 |  \\  \nonumber
&+& \hat{T}_{2;\alpha \beta}   \Gamma_{2;nm} (j, \veck) |n,Y_j,\alpha,1 \rangle \langle m,Y_{j+1},\beta,2 |  \\ 
&+& \hat{T}_{3;\alpha \beta}   \Gamma_{3;nm} (j,\veck) |n,Y_j,\alpha,1 \rangle \langle m,Y_{j-1},\beta,2 |   \bigg],
\eea 
where 
\bea
\Gamma_{1; nm} (j, \veck)&=& F_{nm} \bigg( \frac{\vecq_{1} l_{B}}{\sqrt{2}} \bigg) e^{-\frac{4 \pi \imath p}{q} j -\imath \frac{2}{\sqrt{3}} k_1}, \\
\Gamma_{2; nm} (j,\veck)&=& F_{nm} \bigg( \frac{\vecq _2 l_{B}}{\sqrt{2}} \bigg) e^{\frac{\pi \imath p}{q} (2 j+1)+ \frac{\imath}{\sqrt{3}} k_1 + \imath k_2 },\\
\Gamma_{2; nm} (j,\veck) &=& F_{nm} \bigg( \frac{\vecq_3 l_{B}}{\sqrt{2}} \bigg) e^{\frac{\pi \imath p}{q} (2 j-1)+ \frac{\imath}{\sqrt{3}} k_1 - \imath k_2 },
\eea
where $\veck =(k_1,k_2)$ is defined in units of $\Delta = \sqrt{3}/2 k_{\theta} l_{B}^2 = 3 p/ (q a_{M})$ and the magnetic BZ (mBZ) is defined by  $k_1 \in [0, 6 \pi/(\sqrt{3}\phi) ) $ and $ k_2 \in [ 0, 2 \pi/q )$ in units of $\Delta $.

The LL form factors $F_{n,n'} (x)$, are given by
\begin{equation}  
F_{n,n'}(\vecq) = \bigg\{ \begin{array}{cc}
\sqrt{\frac{n'!}{n!}} \big[ \frac{ \imath q^{\star}}{\sqrt{2}} \big]^{n-n'} 
L_{n'}^{n-n'}(\frac{q^{2}}{2}) e^{-q^{2}/4} & n \geq  n',\\
\sqrt{\frac{n!}{n'!}} \big[ \frac{\imath q}{\sqrt{2}} \big]^{n'-n}
L_{n}^{n'-n}(\frac{q^{2}}{2}) e^{-q^{2}/4} & n' > n ,
\end{array}
\end{equation}
where, $ q = q_x + \imath q_{y} $ with $q = q_x^2 + q_y^2$ and $L_{n'}^{n-n'} $, are the associated Laguerre polynomials. 

\subsection{Appendix B : Topological Properties of tTLG HM butterfly}

Here, we review our procedure for calculating the topological properties. The Bloch function for the $\lambda_i^{th}$ HM bands in tTLG can be expressed as,
\begin{widetext}
\be
\label{wavefunction}
u_{\lambda_i} (\veck) = \sum_{m =1,n}^{q,N} g_{\lambda_i,n,m} (\veck) \sum_{l} \phi_{n} (x - k_y l_{B}^2- \Delta (m + l q))  \exp(\imath \frac{\Delta y}{l_{B}^2} (m+l q)) \exp(-\imath k_x (x- \Delta m - q \Delta l)),
\ee
\end{widetext}
where $\phi_n$ corresponds to the Harmonic oscillator wavefunction and $g_{\lambda_i,m,n} (\veck)$'s are obtained numerically. The Chern number of the bands is calculated using the lattice gauge theory method introduced in Ref.~\onlinecite{Chernnumbercalculation}. For the case of Eq.~\ref{wavefunction}, there are two contributions to the Chern number, one associated with the lattice eigenvectors $g_{\lambda_i,m,n} (\veck)$, the lattice Chern number $\tilde{C}_{\lambda}$, while the other is associated with the band-folded LL wavefunctions $\phi_n$.

The lattice Chern number for M-band multiplet with collection of bands with indices, $\lambda_{M} = (\lambda_1,\lambda_2, \dots, \lambda_M) $ and M-mulitple Bloch wavefunctions $(u_{\lambda_1}, u_{\lambda_2}, \dots, u_{\lambda_M})$ we calculated the Bloch band Chern number  $\tilde{C}_{\lambda_{M}}$ of the multiplet
\be
\tilde{C}_{\lambda} = \frac{1}{2 \pi \imath}  \sum_{\veck_{i}} \prod_{\square} \det \bigg[\frac{ G_{\lambda_M} (\veck_{i}) G_{\lambda_M} (\veck_{i} + \hat{\mu}) }{\det |G_{\lambda_M} (\veck_{i}) G_{\lambda_M} (\veck_{i} + \hat{\mu})|}\bigg],
\ee
where $G_{\lambda_M} (\veck_{i}) = (g_{\lambda_1}(\veck_i), g_{\lambda_2}(\veck_i), \dots, g_{\lambda_1}(\veck_M)) $ is the $N_{d} \times M$ matrix composed of the amplitude of the Bloch band wavefunction. Since our basis set comprises of LLs, each band folded LL carries a Chern number $1/q$. The total Chern number is the sum of the lattice Chern number and LL contribution associated to the multiplet,
\be 
C_{\lambda}  = \tilde{C}_{\lambda} + \frac{M}{q}.
\ee	
	
\subsection{Appendix C: Wannier diagrams}	
\begin{figure}
	\begin{center}
		\includegraphics[width=0.49\textwidth,clip]{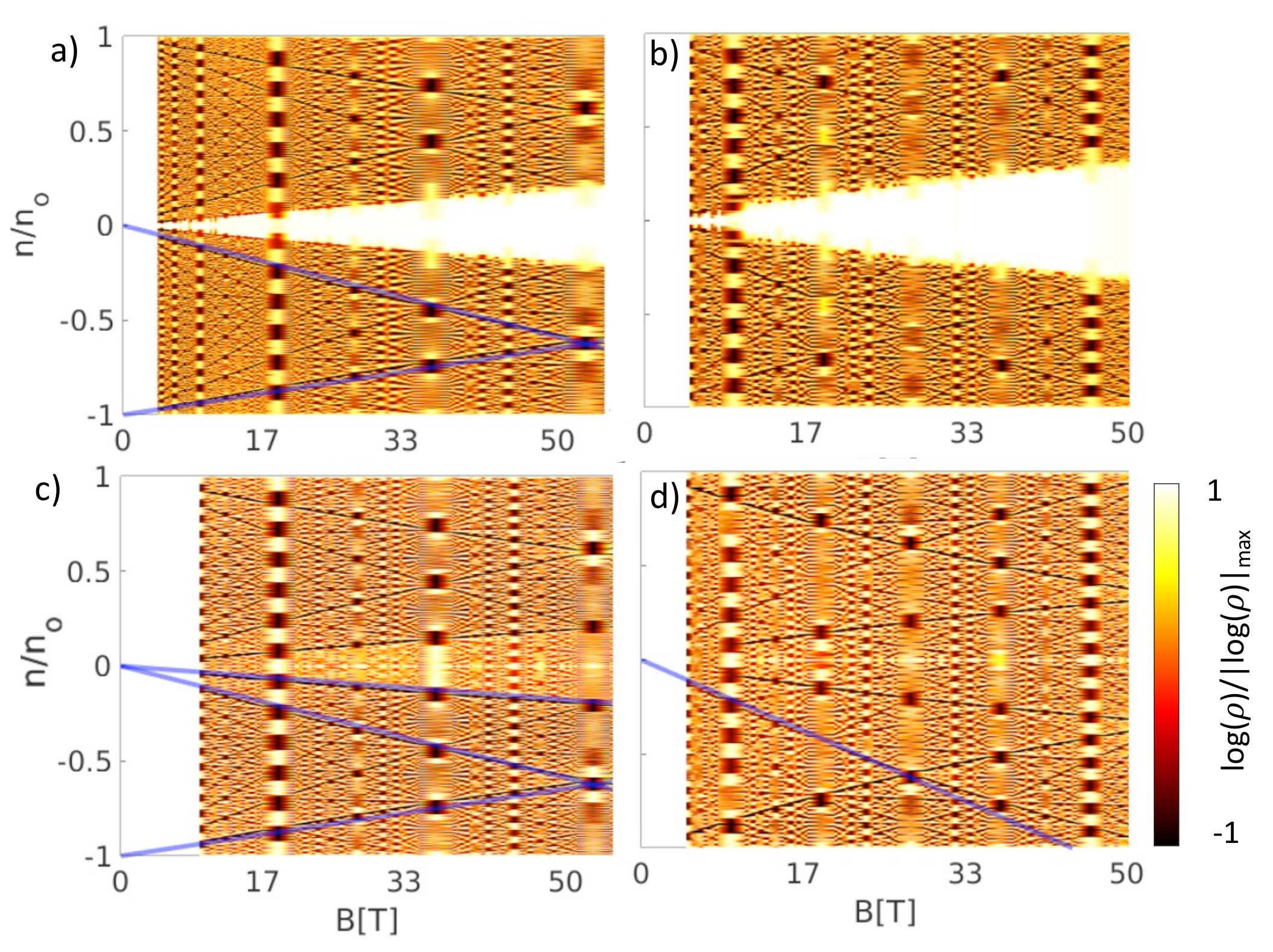}
		\caption{Wannier plots for $\theta =2^{\circ} $ a) and c) and $\theta =1.51^{\circ}$ b) and c), with displacement field $ \Delta_{\perp} =0, 10$ meV, respectively. The straight blue lines are drawn from the Streda formula. We have only shown them for hole-HM bands. The sharp density of state features at the charge neutrality point is associated with the LL of odd parity Dirac band of tTLG. As the gate voltage is applied, these sharp features in the density of states disappear due to mixing with Hofstadter energy bands. The absence of a white gap in the wedge-shaped region at the charge neutrality point indicates this.}	
		\label{fig:Wannierplots}
	\end{center}
\end{figure}	
	
Fig~\ref{fig:Wannierplots} a) and b) show the Wannier plots in the absence of any displacement field, where mirror symmetry is exact, for $\theta = 2^{\circ} $and $\theta = 1.51^{\circ}$, respectively. The slope of the straight lines corresponds to the Hall conductivity as indicated by Streda's formula, $\bar{n} = \sigma\bar{\phi} +c$, where $\bar{n} = n/n_{0}$, refers to the normalized number density ($n_{0}$ is the total density that includes the spin and valley degeneracy), and $\bar{\phi}$, refers to the tight binding alpha $\bar{\phi}=1/(6\phi)$, $\sigma$ denotes the Hall conductivity, and $c$ denotes the intercept. The blue lines guide the eye and indicate Streda's formula applied to the HM pattern of tTLG. The high density of states in the wedge-shaped region Fig~\ref{fig:Wannierplots} a) and b) is due to the simultaneous presence of the odd parity zeroth LL.  

Fig~\ref{fig:Wannierplots} c) and d) show the Wannier plots for $\Delta_{\perp} =10$ meV for $\theta = 2^{\circ} $and $\theta = 1.51^{\circ}$, respectively. The wedge shape region at charge neutrality has reduced intensity, indicating a gap opening in the presence of a displacement field. Additionally, the odd parity Landau bands become weakly dispersive due to mixing with the even parity tBLG energy bands. A displacement field can tune this topological transition; it can be identified in the Wannier plots as a more prominent downward sloping $-2(2) $ lines on the electron(hole) regions at the magic angle in Fig.~\ref{fig:Wannierplots} d).

\subsection{Appendix D: Landauer-Buttiker theory}

Consider a quantum parity Hall phase with $m$-right moving and $n$-left moving channels protected by mirror symmetry in a Hall bar geometry setup. The current into the $i^{th}$-lead $I_{i}$ can be expressed as
\begin{equation} 
I_{i} = \frac{e^2}{h}\sum_j (T_{ij} V_j -T_{ji} V_i),
\end{equation}  
where $T_{ij} $ is the transmission probability of the current from the $j^{th}$ lead to the $i^{th}$ and $V_i$ is the potential associated with the $i^{th}$ lead~\cite{PhysRevB.38.9375}.
We assume perfect contacts i.e. $T_{ij} =1$, and that the current is applied to the $1^{st}$-lead and drained from the $4^{th}$ lead. The rest are floating contacts and act as voltage probes giving $I_2=I_3=I_5 = I_6 =0$. We can write $I_1 = -I_4 = I$ because of the charge conservation. The voltages in terms of the current can be determined by solving the linear system of equations,
\bea 
V_1 &=& - \frac{h}{e^2} \bigg( \frac{n^2}{m^3 + n^3} \bigg) I, \\
V_{2} &=& \frac{h}{e^2} \bigg( \frac{m-n}{m^2-mn+n^2}  \bigg) I,\\
V_3 &=& \frac{h}{e^2} \bigg( \frac{m^2 + mn - n^2}{m^3 + n^3} \bigg) I, \\
V_4 &=& \frac{h}{e^2} \bigg( \frac{m}{m^2 - mn + n^2} \bigg) I,\\
V_5 &=& \frac{h}{e^2} \bigg( \frac{mn}{m^3 + n^3} \bigg) I,
\eea
with $V_6 =0$. The resistance $R_{ij,kl}$ is defined as the ratio of the voltage to the current measured between the $k^{th}$ and the $j^{th}$, with current applied from the $i^{th}$ to the $j^{th}$ lead gives,
\be 
R_{14,26} = R_{14,35} =  \frac{h}{e^2} \bigg( \frac{m-n}{m^2 -mn + n^2} \bigg) ,
\ee 
while 
\be 
R_{14,32} = R_{14,56} = \frac{h}{e^2} \bigg( \frac{mn}{m^3 + n^3} \bigg),
\ee 
and 
\be 
R_{14,41} = \frac{h}{e^2} \bigg( \frac{m^2+mn+n^2}{m^3 + n^3} \bigg) .
\ee

\end{document}